\documentclass[twocolumn,prx,superscriptaddress,longbibliography]{revtex4-1}
\usepackage{amssymb}
\usepackage{amsmath}
\usepackage{color}
\usepackage[colorlinks=true,urlcolor=blue,breaklinks,citecolor=blue]{hyperref}
\usepackage{graphicx}
\usepackage{braket}
\usepackage{subfigure}
\usepackage{xcolor}
\usepackage{dcolumn,bm}
\usepackage{cleveref}
\usepackage{float}
\usepackage{here}
\usepackage{hyperref}

\usepackage[final]{changes}

\newcommand{\CSF}{C_\text{SF}}
\newcommand{\CDW}{C_\text{DW}}
\newcommand{\Cs}{C_\text{HI}}
\newcommand{\chimax}{\chi_\text{max}}
\newcommand{\nmax}{n_\text{max}}

\begin{document}
\title{Unsupervised phase discovery with deep anomaly detection}

\author{Korbinian Kottmann}
\author{Patrick Huembeli}
\affiliation{ICFO - Institut de Ciencies Fotoniques, The Barcelona Institute of Science and Technology, Av. Carl Friedrich Gauss 3, 08860 Castelldefels (Barcelona), Spain}
\author{Maciej Lewenstein}
\author{Antonio Ac\'in}
\affiliation{ICFO - Institut de Ciencies Fotoniques, The Barcelona Institute of Science and Technology, Av. Carl Friedrich Gauss 3, 08860 Castelldefels (Barcelona), Spain}
\affiliation{ICREA, Pg. Lluís Companys 23, 08010 Barcelona, Spain}
\begin{abstract}
We demonstrate how to explore phase diagrams with automated and unsupervised machine learning to find regions of interest for possible new phases. In contrast to supervised learning, where data is classified using predetermined labels, we here perform anomaly detection, where the task is to differentiate a normal data set, composed of one or several classes, from anomalous data.  As a paradigmatic example, we explore the phase diagram of the extended Bose Hubbard model in one dimension at exact integer filling and employ deep neural networks to determine the entire phase diagram  in a completely unsupervised and automated fashion.  As input data for learning, we first use the entanglement spectra and central tensors derived from tensor-networks algorithms for ground-state computation  and later we extend our method and use experimentally accessible data such as low-order correlation functions as inputs. Our method allows us to reveal a phase-separated region between supersolid and superfluid parts with unexpected properties, which appears in the system in addition to the standard superfluid, Mott insulator, Haldane-insulating,  and density wave phases.
\end{abstract}

	\maketitle
	
	\paragraph*{Introduction}
	
	Recent developments in machine learning (ML) have revolutionized the way how we can process and find correlations in complex data. These developments have impacted the physical sciences with a wide variety of applications \cite{Carleo2019}. Of particular interest is the classification and discovery of  phase transitions  \cite{wang2016discovering,nieuwenburg2017learning,carrasquilla2017machine,schindler2017probing,liu2017self,wetzel2017unsupervised,chng2017unsupervised,koch-janusz2017mutual, huembeli2018identifying, huembeli2019automated, deng2016exact,zhang2017machine, broecker2017machine, Tsai2019,Shinjo2019, theveniaut2019neural,Dong20182,Kawaki2017}. Recent works concern studies of classical~\cite{wang2016discovering, carrasquilla2017machine}, quantum~\cite{broecker2017machine, huembeli2019automated,Dong20182} and topological phase transitions~\cite{deng2016exact,zhang2017machine}. The methods employed range from deep supervised \cite{carrasquilla2017machine,Dong20182} and unsupervised~\cite{nieuwenburg2017learning, liu2017self} to shallow unsupervised ML algorithms~\cite{wang2016discovering,wetzel2017unsupervised,chng2017unsupervised, nussinov2016inference}. The input of the ML algorithms can vary from classical spin values~\cite{carrasquilla2017machine}, local observables~\cite{wetzel2017unsupervised, chng2017unsupervised}, correlation functions~\cite{broecker2017machine}, entanglement spectra~\cite{nieuwenburg2017learning,schindler2017probing, Tsai2019,Shinjo2019} to the full state vector \cite{huembeli2019automated,theveniaut2019neural}.
	At the same time, the development of the density matrix renormalization algorithm \cite{White1992,White1993} and its reformulation from a quantum information perspective in terms of tensor networks \cite{2011_SCHOLLWOCK,Orus2013} allows one to study large  quantum many-body systems approaching the thermodynamic limit.
	
In this work, we demonstrate how to map out a phase diagram of a quantum many-body system to identify regions of interest for possible new phases using automated and unsupervised machine learning based on anomaly detection \cite{chalapathy2019deep, kwon2017survey,borghesi2019anomaly}.  This approach is particularly useful when one is confronted with sufficient data from known classes of states and little or no data from unknown classes. 


 Compared to previous unsupervised attempts in \cite{nieuwenburg2017learning, liu2017self,wang2016discovering,wetzel2017unsupervised,chng2017unsupervised}, this method needs only one or few training iterations and has better generalization properties from employing deep neural networks \cite{kawaguchi2017generalization, valle2018deep}. This allows for efficient fully automatized phase discovery in the spirit of self-driving laboratories \cite{Hase2019}, where artificial intelligence augments experimentation platforms to enable fully autonomous experimentation. Intuitively, the method explores the phase diagram until an abrupt change, an anomaly, is detected, singling out the presence of a phase transition. The intuition is similar to the approach introduced in~\cite{Zanardi2006}, where the authors proposed to detect quantum phase transitions by looking at the overlap between neighbouring ground states in the phase diagram. Here, the machine is used to detect these anomalies. Moreover, as we explain next, it does it from scalable data.

In principle, there are many possible choices as input data for training our method, including the full state vector. To improve scalability and reach large system sizes, we propose to use quantities that arise naturally in the state description and do not require complete state information. For instance, we obtain ground states with tensor networks, from which we use the tensors themselves or the entanglement spectrum (ES) as input data. These quantities arise naturally from the state description without further processing and contain crucial information about the phase, like ES for example \cite{Deng2011,Shinjo2019,Tsai2019}. We stress, however,  that the choice of preferred quantities to be used for ML may in general vary and depend on the simulation method. In fact, we show that our method also works well with physical data accessible in  experiments such as low-order correlation functions.

As a benchmark, we apply our method to the extended Bose Hubbard model in one dimension at exact integer filling. Its phase diagram is very rich and therefore provides a very good test to showcase our method. We are able to  determine the 
entire phase diagram  in a completely unsupervised and automated fashion.  Importantly, our results point out the existence of a supersolid state that appears in the system in addition to the standard superfluid, Mott insulator, Haldane-insulating, and  density-wave phases.

\paragraph*{Anomaly Detection Method}

In this work, we apply deep neural network autoencoders for anomaly detection \cite{borghesi2019anomaly}.
An autoencoder (AE) is a type of neural network that consists of two parts. The encoder part takes the $D$-dimensional input data point $x$ and maps it to a $k$ dimensional latent variable $z$ (typically $k<D$) via a parametrized function $z = f_{\phi}(x)$. The decoder part takes the latent variable $z$ and maps it back to $\bar{x} = g_{\theta}(z)$. The parameters $\phi$ and $\theta$ are trained via the minimization of a loss function $L(x,\bar{x})$ that measures the dissimilarity of the input $x$ and the output $\bar{x}$. The aim of the training is that the input is identical to the output for the whole training data set $\{x \}$. Heuristically, we find that the \textit{mean-square error} $L(x,\bar{x}) = \sum_v |x_v - \bar{x}_v|^2/D$ suffices for this endeavour and provides good results.

\begin{figure}[tb]
\begin{center}
\includegraphics[width=.3\textwidth]{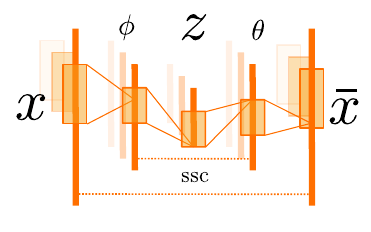}
\end{center}
\caption{\label{fig:AE} Schematic one-dimensional convolutional neural network autoencoder with symmetric shortcut connections (SSC) that connect latent layers of the same dimension directly via addition, thereby improving the model performance \cite{Dong2016,Dong2016_2}. Shaded boxes in the background indicate schematically the convolutional nature of the layers. We illustrate how the input data $x$ gets encoded into the lower dimensional latent vector $z$ and decoded again to $\bar{x}$.}
\end{figure}

The idea of this anomaly detection scheme is that for each state $\ket{\psi}$ we take corresponding data $x$, such as for instance its ES or low order correlation functions. That data has characteristic features that the AE learns to encode into the latent variable $z$ at the bottleneck \cite{Iten2020},
from which it is ideally able to reconstruct the original input. The loss $L$ directly indicates the success of this endeavour, which we improve by employing symmetric shortcut connections (SSC, see \cref{fig:AE}), inspired from \cite{Dong2016,Dong2016_2} to typical losses $<5\%$. Now, the intuition is that, 
when confronted with data from unknown phases, the AE is unable to encode and decode $x$. This leads to a higher loss, from which we deduce that the states do not belong to the same phase as the ones used to train the AE.

Deep learning architectures are known to generalize well \cite{kawaguchi2017generalization, valle2018deep}, such that it suffices to train in a small region of the parameter space. Compared to known supervised deep learning methods this anomaly detection scheme does not rely on labeled data. We choose training data from one or several regions of the phase diagram, and ask how the loss of a test data point from any region of the phase diagram compares to the loss of these training points. As we show later, this can be performed with no a priori knowledge and in a completely unsupervised manner. The computationally most expensive step is the training and with our method it has to be performed only once to map the whole phase diagram, as opposed to multiple trainings like in \cite{nieuwenburg2017learning, liu2017self}. Furthermore, it does not require a full description of the physical states in contrast to \cite{Zanardi2006}, where full contraction is necessary. Thus, for higher dimensional systems, \cite{Zanardi2006} is infeasible as contraction is known to be generally inefficient for 2d tensor network states (commonly referred to as PEPS, see \cite{Orus2013}).

The specific architecture in use consists of two 1d-convolutional encoding and decoding layers with SSCs (Figure \ref{fig:AE}), implemented in TensorFlow \cite{tensorflow2015-whitepaper}.
To ensure the reproducibility of our results, we made the source code available under an open source license~\cite{githubrep}.

\paragraph*{Simulation Method}

We calculate the ground states by means of the Density Matrix Renormalization Group algorithm (DMRG) in terms of Tensor Networks, i.e. Matrix Product States (MPS) \cite{2011_SCHOLLWOCK,Orus2013}. A general multipartite state of $L$ parties with local dimension $d$ $\ket{\Psi} = \sum_{\bm{\sigma}} c_{\bm{\sigma}} \ket{\bm{\sigma}}$, where $\bm{\sigma}=\sigma_1\ldots\sigma_ L$ is the vector of local indices $\sigma_i = 1,\ldots,d$, can always be decomposed into products of tensors with the aid of the singular value decomposition. We use the convention of Vidal \cite{Vidal2003}, and write our ground state in the MPS form

\begin{multline}
\label{eq:ansatz_mps}
\ket{\Psi} = \sum_{\bm{\sigma}} \Gamma^{\sigma_1} \Lambda^{[1]} \cdots \Lambda^{[i-1]} \Gamma^{\sigma_i} \Lambda^{[i]} \cdots  \\
\Lambda^{[L-1]} \Gamma^{\sigma_{L}} \ket{\sigma_1 \ldots \sigma_i \ldots \sigma_{L}}.
\end{multline}

At site $i$, $\{\Gamma^{\sigma_i}\}$ is a set of $d$ matrices and $\Lambda^{[i]}$  the diagonal singular value matrix of a bipartition of the chain between site $i$ and $i+1$, i.e. the Schmidt values (see \cite{2011_SCHOLLWOCK}). One then approximates the exact ground state by keeping only the $\chimax$ largest Schmidt values for each partition, where $\chimax$ is known as the bond dimension. This is the best approximation of the full state in terms of the Frobenius norm and enables us to handle big system sizes. Eq. (\ref{eq:ansatz_mps}) corresponds to finite length and open boundary conditions. Here, we use the version formulated in the thermodynamic limit for infinite MPS (iMPS) \cite{Vidal2006,McCulloch2007,Orus2008}. In this case, instead of a finite chain, we are effectively operating in the thermodynamic limit and have a finite but repeating unit cell of length $L_\infty$.

We use the Schmidt values $\Lambda^{[i]}$ as our input data $x$ to explore the phase diagram and ambiguously refer to it as ES. Our numerical results support the functionality of using this anomaly detection scheme with ES as we get near-constant losses for states of the training region and significantly higher losses for unknown phases. The method generalizes well with similar losses for states inside and outside the training region. As we will see below, the method works even well for transitions of Berezinskii-Kosterlitz-Thouless (BKT) type, where the exact transition point is hard to determine in terms of observable correlation functions, and symmetry protected topological (i.e. global) order.

\paragraph*{Hamiltonian}

We test our method on the extended Bose-Hubbard Model
\begin{multline}
H = -t \sum_i \left( b^\dagger_i b_{i+1} + b^\dagger_{i+1} b_i \right) \\
+ \frac{U}{2} \sum_i n_i(n_i -1) + V \sum_i n_i n_{i+1},
\end{multline}
with nearest neighbour interaction on a one dimensional chain. It serves as a highly non-trivial test ground with its rich phase diagram that, beside a critical superfluid  and two insulating phases, admits a symmetry protected topologically ordered phase at commensurate fillings  \cite{Rossini2012,1997_Kuehner,Kuehner1999,Mishra2009,Urba2006,
Ejima2014,Cazalilla2011,Batrouni2006,Deng2011,Berg2008}. Here, $n_i = b^\dagger_i b_i$ is the number operator for Bosons defined by $[b_i,b_j^\dagger] = \delta_{ij}$. Typically, we are interested in varying the on-site interaction $U$ and nearest-neighbour interaction $V$ and fix the hopping term $t=1$.  We explicitly enforce filling $\bar{n} :=\sum_i \braket{n_i}/L_\infty = 1$ by employing $U(1)$ symmetric tensors \cite{Silvi2017}, which we implement using the open source library TeNPy \cite{tenpy} (easily readable code accessible in \cite{githubrep}).

\begin{figure}[tb]
\begin{centering}
	\includegraphics[width=.49\textwidth]{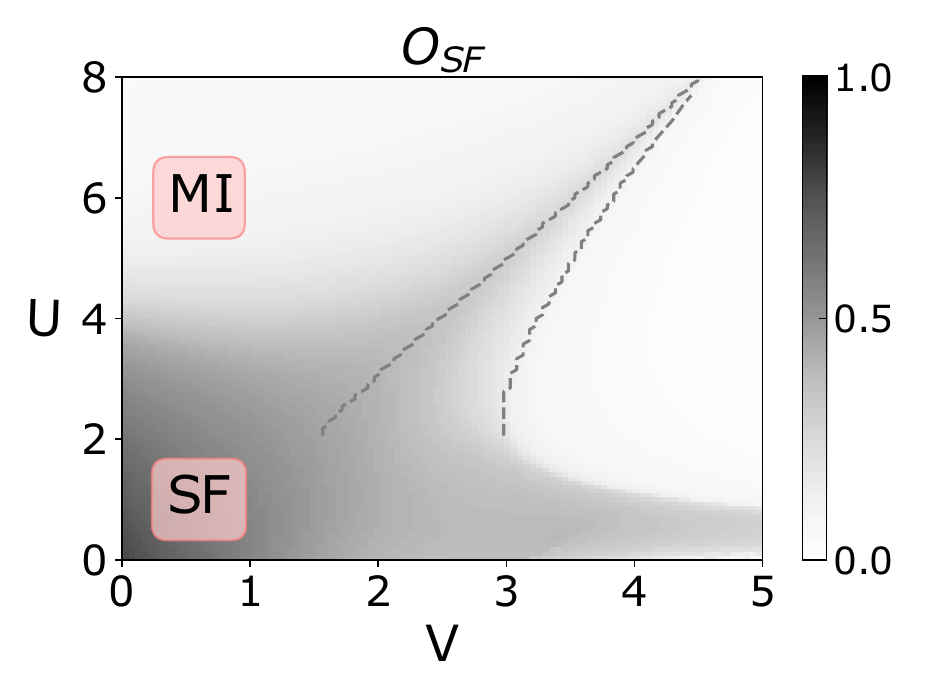}
	\includegraphics[width=.49\textwidth]{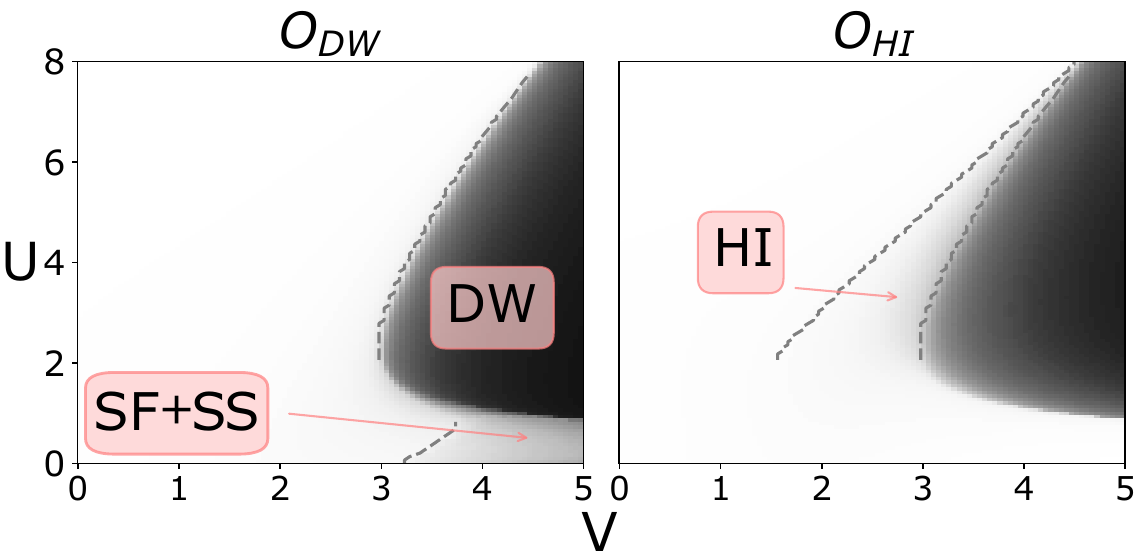}
\end{centering}
\caption{\label{fig:phase_diagram} Extended BH phase diagram with five distinct phases obtained by the correlators \cref{eq:CSF,eq:CDW,eq:Cs}. MI: Mott Insulator, SF: Super Fluid, SS: Super Solid, DW: Density Wave, HI: Haldane Insulator. The dashed lines indicate the transition points observed from diverging correlation lengths between MI-HI-DW and non-zero $\mathcal{S}$ in \cref{eq:solidity} between SF and SF+SS.}
\end{figure}

One way to physically classify these phases is to look at the correlators
\begin{align}
\CSF(i,j) &= \braket{b^\dagger_i b_j} \label{eq:CSF} \\
\CDW(i,j) &= \braket{\delta n_i (-1)^{|i-j|} \delta n_j} \label{eq:CDW} \\
\Cs(i,j) &= \braket{\delta n_i \exp\left( -i \pi \sum_{i \leq l \leq j-1} \delta n_l\right) \delta n_j}\label{eq:Cs}
\end{align}
with $\delta n_i = n_i - \bar{n}$. $\CSF$ discriminates the Mott-insulating (MI) phase and the superfluid (SF) phase, where it decays exponentially and with a power-law, respectively. The correlators for density-wave (DW) and Haldane-insulating (HI) phases decay to a constant value in the respective phases. More details about the characterization of the system can be found in \cite{Rossini2012} and the supplementary material (SM). The non-local string term in \cref{eq:Cs} is characteristic of topological order, where the translational symmetry remains protected with a transition in the Luttinger liquid universality class from MI and gets broken with a transition in the Ising universality class to DW \cite{Berg2008}. We visualize the phase diagram by computing $O_{\bullet} = \sum_{i,j} C_{\bullet}(i,j)/L_\infty^2$ in \cref{fig:phase_diagram} in the thermodynamic limit for a repeating unit cell of $L_\infty=64$ sites with a maximum bond dimension $\chimax = 100$ and assuming a maximum occupation number $\nmax=3$, which results in a local dimension $d=\nmax+1=4$. We use data from these states, obtained with these parameters throughout the rest of the following analysis.

\begin{figure}[tb]
\begin{center}
\includegraphics[width=.49\textwidth]{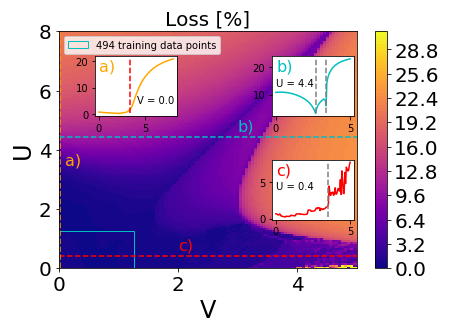}
\end{center}
\caption{2D loss map of the AE after training near the parameter space origin (blue square frame). The insets a), b) and c) show the loss along the dashed lines. Vertical green dashed line in inset a) indicates critical $U_c = 3.33$ \cite{Kuehner1999}. Vertical grey dashed lines in inset b) and c) are the transitions from \cref{fig:phase_diagram}. The phase boundaries are determined by a rise in loss (inset a) and c)). The anomalous regions are already well-separated by decreasing losses because of the critical behaviour at the phase boundaries (inset b)), which share similarities with the critical SF phase. Higher loss indicates that this region is more different from the training region in the blue square, lower loss indicates similarity.}
\label{fig:2D_pbc_ES_SF}
\end{figure}

\paragraph*{Numerical Results}

Assuming no a priori knowledge, we start by training with data points at the origin of the parameter space $(U,V) \in [0,1.3]^2$, which in our case accounts to training in SF. By testing with data points from the whole phase diagram we can clearly see the boundaries to all other phases from SF in \cref{fig:2D_pbc_ES_SF}. The BKT transition between SF and MI is matched by an abrupt rise in loss (\cref{fig:2D_pbc_ES_SF}, inset a)). In this particular case, we can already determine the different phases inside the anomalous region due to their different loss levels and the appearance of two valleys at the phase boundaries between MI, HI and SF (\cref{fig:2D_pbc_ES_SF}, inset b)). Physically, we can explain these valleys by the criticality of these Luttinger and Ising type transitions, which lead to a slowly decaying ES at the boundary, just like in the critical SF phase. 

\begin{figure}
\begin{center}
\includegraphics[width=.49\textwidth]{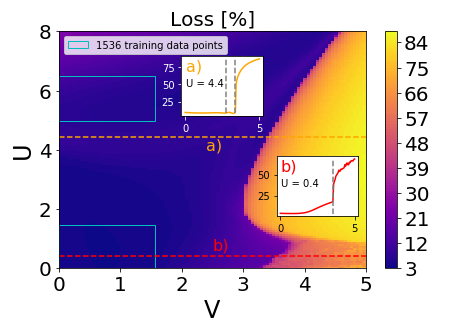}
\end{center}
\caption{2D loss map of the AE after training in the two blue square frames in the SF and the MI phase. The insets a) and b) show the loss along the dashed lines. Instead of the ES, we use the physically accessible correlator $\CSF$ as input data. The HI is not recognized as this correlator does not contain information about the topological order of this phase.}
\label{fig:2D_pbc_cor_MI-SF}
\end{figure}

It is not necessarily always the case that one can differentiate the different phases inside the high-loss anomalous region. Thus, as a systematic approach, we propose picking homogeneous and high contrast anomalous regions after the initial training. Here, we already mapped out the whole phase diagram after the first training iteration, so we leave a possible continuation in $(U,V) \in [4,4.8]\times[2,4]$ to the SM. 
This method is not tailored to ES as input data. To show this, we use on one hand tensors from the MPS as input data in the SM. On the other hand we use experimentally accessible correlators. In \cref{fig:2D_pbc_cor_MI-SF}, instead of unprocessed data from simulation, we calculate $\{\CSF(i,j)\}_{i,j=1}^{64}$ and train in MI and SF simultaneously. We interpret rows as color channels for 1d convolution. Because $\CSF$ does not contain any information about the topological order in HI, the method does not recognize this region as we would expect (\cref{fig:2D_pbc_cor_MI-SF}, inset a)). Overall, the boundaries match perfectly with a sharp increase onto a plateau at the transition points. This opens the possibility to use physical observables from experiment with the caveat of requiring physical knowledge a priori.

\begin{figure}
\begin{center}
\includegraphics[width=.49\textwidth]{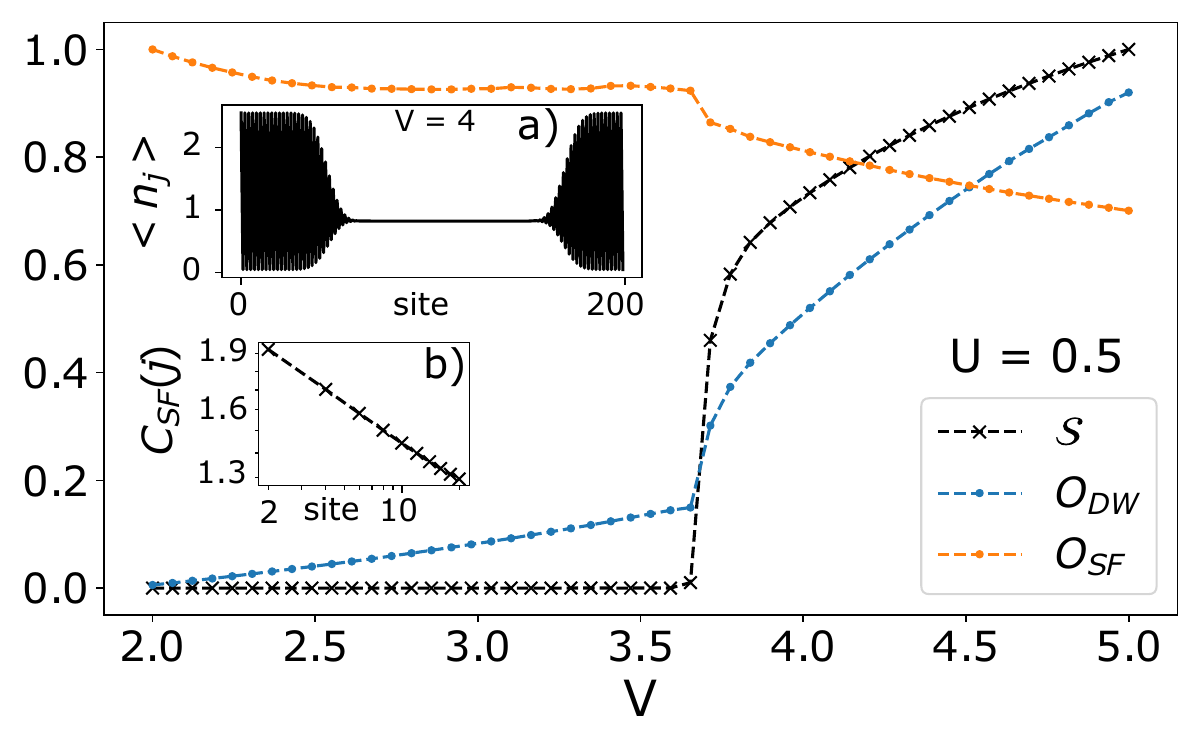}
\end{center}
\caption{Transition from SF to phase separated SF + SS at fixed U=0.5. The solid long-range order emerges while SF correlations sustain. Inset a) shows the phase separation in the density $\braket{n_j}$ for a state at $(U,V)=(0.5,4)$ with $L_\infty=200$, $d=6$ and $\chimax = 500$. Inset b) shows the power-law decay of $\CSF(0,j)$ in the solid part via doubly logarithmic plot of every second value, confirming supersolidity.}
\label{fig:SS_master}
\end{figure}

By close inspection of \cref{fig:2D_pbc_ES_SF,fig:2D_pbc_cor_MI-SF}, we see a region with noticeable contrast for small $U$ and large $V$, indicating the presence of a separate phase. This is interesting because, initially, we did not expect to find a fifth phase in the diagram. Upon further physical investigation, we find a phase-separated state between SF and supersolid (SF + SS). Supersolidity in this model has been studied in previous literature for incommensurate fillings \cite{1997_Kuehner,Kuehner1999,Mishra2009,Kawaki2017} and was claimed to be found for filling 1 in \cite{Deng2011} without further discussion. The phase separation that we find here is new and has not been studied before to the best of our knowledge. In order to physically show the transition, we compute the Fourier transform of the local density $\tilde{n}(k) = \sum_j \braket{n_j} e^{-ikj}/L_\infty$ and detect long-range solid order by looking at
\begin{equation}
\label{eq:solidity}
\mathcal{S} := \max_{k\neq0} \left|\tilde{n}(k)\right|^2
\end{equation}
in \cref{fig:SS_master} \cite{Chen2017}. Additionally, we find non-zero $O_{DW}$ and $O_{SF}$, showing both superfluid and crystalline behavior.
For higher numerical accuracy and better illustration of the correlator decay, we compute a larger state for $L_\infty=200$, $d=6$ and $\chimax = 500$ at $(U,V)=(0.5,4)$ and see both crystalline and superfluid regions in the density profile $\braket{n_j}$, \cref{fig:SS_master} inset a). To confirm supersolidity of the crystalline part we show that $C_{SF}$ decays with a power-law in that region, see \cref{fig:SS_master} inset b). 

\begin{figure}
\begin{center}
\includegraphics[width=.49\textwidth]{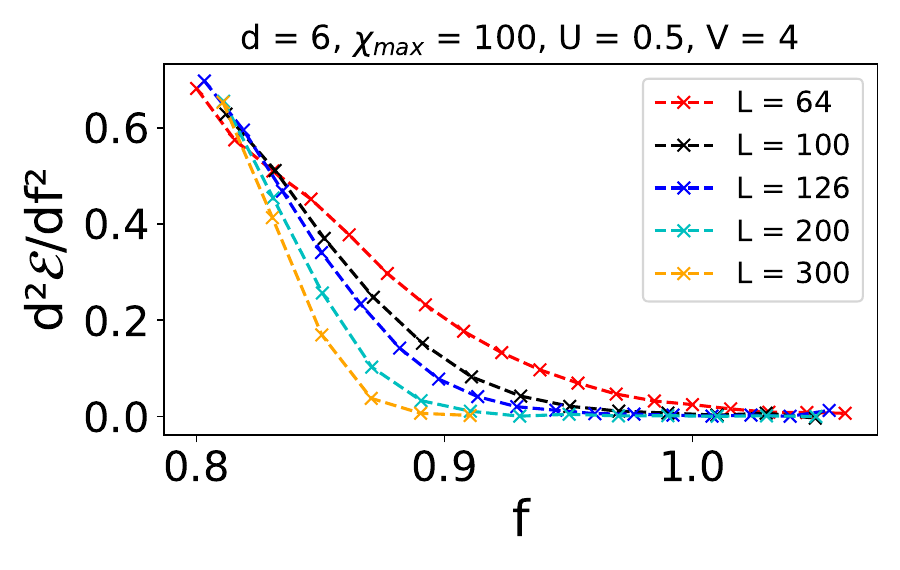}
\end{center}
\caption{Finite size scaling of the vanishing second derivative of the ground state energy per site $\mathcal{E}$ with respect to the filling $f$. This shows that the system becomes mechanically unstable, leading to phase separation as depicted in \cref{fig:SS_master} for $f=1$.}
\label{fig:OBC_d2_f_n-max-5_raw}
\end{figure}

This phase separation occurs as the system becomes mechanically unstable. We can see this as the second derivative of the ground state energy per site $\mathcal{E} = E/L$ with respect to the filling $f$ vanishes. We perform finite size scaling with open boundary conditions to show this in \cref{fig:OBC_d2_f_n-max-5_raw}. There, we target equidistant discrete fillings $f_i  = N_i/L$ for $N_i \in [0.8 L,1.1 L]$ and compute the finite difference derivative $d^2\mathcal{E}/df^2 = \left(\mathcal{E}(f_{i-1}) - 2 \mathcal{E}(f_i) + \mathcal{E}(f_{i+1})\right)/(f_i-f_{i-1})^2$. The detection of this new phase demonstrates the power of our approach and we leave further physical investigation to future work.

\paragraph*{Conclusion}
We have shown an unsupervised method to map out the phase diagram of a complex quantum many-body system that could possibly be performed fully data driven and without physical a priori knowledge such as the construction of an order parameter. By using tensor networks we can reliably compute ground states of many-body systems in the thermodynamic limit and at the same time extract the desired data without further processing. Entanglement spectra  and central tensors serve as natural quantities in this context, but the method also proved successful for physical observables like $\braket{b^\dagger_i b_j}$ correlators. Hence, this method can be applied in both purely computational platforms like self-driving laboratories as well as experimental setups.

\begin{acknowledgements}
We thank E. Tirrito, D. Gonzalez-Cuadra, A.Dauphin, G. Astrakharchik and P. Massignan for helpful insights and discussions. This project has received funding from the European Union’s Horizon 2020 research and innovation programme under the Marie Sklodowska-Curie grant agreement No 665884 (P.H.) and 713729 (K.K.). We acknowledge the Spanish Ministry MINECO (National Plan 15 Grant: FISICATEAMO No. FIS2016-79508-P, TRANQI, SEVERO OCHOA No.SEV-2015-0522), European Social Fund, Fundacio Cellex and Mir-Puig, Generalitat de Catalunya (AGAUR SGR 1341, SGR1381, QuantumCAT and CERCA/Program), ERC AdGs NOQIA and CERQUTE, the AXA Chair in Quantum Information Science and the National Science Centre, Poland-Symfonia Grant No.2016/20/W/ST4/00314.
\end{acknowledgements}

\bibliography{lit}

\end{document}